\documentclass[11pt]{article}
\usepackage{authblk}
\usepackage{amsmath}	
\usepackage{graphicx}
\usepackage[margin=1in]{geometry}
\usepackage{setspace}
\usepackage{abstract}
\usepackage{float}
\usepackage{amsfonts}
\pdfoutput=1

\title{\vspace{-2cm}Multivariate analysis of Brillouin imaging data by supervised and unsupervised learning}

\author[1]{YuChen Xiang}
\author[1]{Kai Ling C. Seow}
\author[1]{Carl Paterson}
\author[2]{Peter T\"or\"ok*}

\affil[1]{Blackett Laboratory, Department of Physics, Imperial College London, Prince Consort Road, London, SW7 2AZ, UK}
\affil[2]{Division of Physics and Applied Physics, Nanyang Technological University, Singapore}

\affil[*]{Corresponding author: peter.torok@ntu.edu.sg}

\begin{document}
	
	\maketitle
	
\begin{abstract}
	Brillouin imaging relies on the reliable extraction of subtle spectral information from hyperspectral datasets. To date, the mainstream practice has been using line fitting of spectral features to retrieve the average peak shift and linewidth parameters. Good results, however, depend heavily on sufficient SNR and may not be applicable in complex samples that consist of spectral mixtures. In this work, we thus propose the use of various multivariate algorithms that can be used to perform supervised or unsupervised analysis of the hyperspectral data, with which we explore advanced image analysis applications, namely unmixing, classification and segmentation in a phantom and live cells. The resulting images are shown to provide more contrast and detail, and obtained on a timescale $10^2$ faster than fitting. The estimated spectral parameters are consistent with those calculated from pure fitting.
\end{abstract}
\vspace{0.5cm}

\section{Introduction}
Brillouin imaging (BI) is a growing field that has received research interest from various disciplines in recent years. The main attraction of the technique is the ability to probe mechanical properties of structurally complex samples, especially of biological nature with optical resolutions, in a non-contact manner. Experimentally, this was first demonstrated by Koski et al. \cite{BIkoski2005}. The fusion of spectroscopy and imaging has advanced the technique into a powerful, hyperspectral (HS) modality, which has been used to study a wide range of samples, including cells \cite{BRI_Scarcelli2015a}, embryos \cite{Raghunathan2017a}, biofilms \cite{Karampatzakis2017}, human cornea \cite{Gouveia2019}, tumours\cite{Margueritat2019} etc., covered in greater detail in the review by Meng et al. \cite{BRI_Meng2016}. While the development of the instrument is reaching a state of maturity, with ongoing effort focusing on faster, more efficient designs to facilitate \emph{in vivo} applications \cite{Xiang2017,Krug2019,Fiore2019}, more work is still needed on the analysis and interpretation of the hyperspectral datasets \cite{Antonacci2020}. Currently, the most common analysis method is to fit spectral features individually, usually by a single peak in each spectrum to extract the average Brillouin shift and linewidth information, which are then mapped to achieve imaging. This protocol, however, has been shown to rely heavily on a good SNR and even then may still suffer from bias due to noise and complex sample structures \cite{Xiang2020,Mattana2018}. Moreover, using least-squares fitting to form a typical Brillouin image, which may easily comprise up to $10^4$ pixels (spectra), quickly becomes a time-consuming procedure. The use of spectral phasor and moment\cite{Elsayad2019,Ioretto2019}, have been proposed to correct the inaccuracy of fitting a single peak to a noisy, multi-component spectral feature, with the former being truly fitting-free and relying on faster computational processes. In other HS imaging applications, multivariate methods of the data cubes have become a major area of interest, where learning algorithms have been demonstrated to rapidly provide enhanced information from the sample, which can be used for real-time diagnosis, at the point-of-care or even during surgical operations \cite{Ortega2019,Grigoroiu2020,Fabelo2018}. To our knowledge, such methods have only been applied once before for BI in a study of amyloid-beta plaques\cite{Palombo2018}, where a semi-supervised variant originally designed for CARS imaging \cite{Masia2013} was used to decompose the raw data into components. We present here an entire range of multivariate algorithms, both unsupervised and supervised, that are applicable to Brillouin hyperspectral data. These methods are first introduced in Section \ref{sec:methods} with comparison of their relative strengths and inner-workings more specific to Brillouin applications. Additionally, we propose a new hybrid algorithm that can be used to include \emph{a priori} information from other imaging modes into an unsupervised model. Multivariate analysis of Brillouin data from both a phantom and 3T3L1 fat cells are presented in Section \ref{sec:results}. Apart from improved image quality and processing speed, the utility of some methods in particular are highlighted as we show how they can be implemented to perform spectral unmixing, segmentation and classification, which are all sought after functions in clinical applications. 

\section{Supervised and unsupervised multivariate algorithms} \label{sec:methods}
In essence, multivariate (MV) analysis derives from the idea that the relationship between different sets of experimental variables may be used to build a overall statistical model \cite{Anderson1985}. In the case of Brillouin imaging, it is based upon the relative comparability of spectra collected from different parts of the sample. The use of MV analysis in other sectors of hyperspectral imaging has already become a full-fledged member of the data analysis toolbox. For example, the analysis of hyperspectral imagery in remote sensing has been practised in the last 20 years, such as in the study of planetary surfaces \cite{Ceamanos2016}. These methods have also achieved wide-spread acceptance in the field of Raman Imaging \cite{Bonnet1992,Hanlon2000,Hassing2011}. In this work we investigate the applicability of these techniques for Brillouin imaging. Note that the list of methods is not intended to be exhaustive, but rather provides an overview of potential applications. In machine learning, algorithms are commonly classified as supervised or unsupervised. In this work, the former definition refers to algorithms which function optimally with some \emph{a priori} knowledge (e.g. the number of constituents in the field of view), whereas the latter is capable of finding underlying patterns with only the input, experimental data. In some cases, however, some supervision of inherently unsupervised algorithms give the best results i.e. a semi-supervised approach. Thus the terminology is only used here to facilitate the discussion in terms of general similarities and differences between methods. All data analysis was implemented in the \emph{Matlab} environment, using the open-source Biodata protocol \cite{DeGussem2009}, modified to include all necessary functions for Brillouin data, including data storage, labelling and pre-processing, which consists of background-subtraction \cite{Lieber2003}, standard-normal variate (SNV) normalisation \cite{statsbook2} and denoising according to Ref.\cite{Xiang2020}.

\subsection*{principal component analysis}
The best example of unsupervised multivariate analysis and perhaps the most common is principal component analysis (PCA), which is based on the original concept by Pearson from 1901 \cite{Pearson1901}. Conceptually all PCA algorithms perform a transformation on the raw data matrix obtained from hyperspectral imaging, following a linear model \cite{Anderson1985}. The raw data matrix of is denoted by $\mathbf{D} \in \mathbb{R}^{N_xN_y \times N_s}$ and consists of a linear stack of spectra sampled by $N_s$ points, acquired at $N_xN_y$ different spatial locations: 
\begin{equation}
\hat{\mathbf{D}}=\mathbf{SP^T}+\hat{\mathbf{N}}
\end{equation}
where the transformation yields $p$ principal components or loadings, denoted by $\mathbf{P}$, with a data dimension of $Ns\times p$. These then correspond to the \emph{endmembers} of the data and their relative abundance is given by the \emph{score} of each component $\mathbf{S}$, with a data dimension of $N_xN_y\times p$. Any residue in the data that is not explained by this transformation (for example, due to noise), is accounted for by the matrix $\mathbf{N}$, which has the same dimension as the original data. PCA reduces the effective dimension of the data cube, which makes it an ideal precursor for subsequent supervised learning \cite{Swets1996} (e.g. discriminant analysis). On the other hand, the vectors of scores can also be rearranged to form images of abundance for chosen components, composite images can also be formed by assigning each with a different colour. A typical implementation of PCA for spectral analysis starts by projecting the largest variation within the raw spectra onto a vector, which is conventionally the first principal component and then more components are defined until all detectable variations have been considered. Statistically, the data have merely been rotated to a new coordinate system, described by the set of principal components, which inherently displays the maximum variance. Since PCA simply represents the same data differently, the transformation itself, which is orthogonal, can always be performed on any dataset, which is one property that differentiates PCA from similar analyses, such as common factor analysis\cite{Velicer1990}. It should also be pointed out that the incorporation of the residual matrix $\mathbf{N}$ coincides well with an additive noise model, which holds true for most Brillouin instruments \cite{Foreman2019} and directly explains the noise resilience of most PCA algorithms. Since only a few components are necessary to describe the statistically significant variance in the data, the noise terms attributed to unwanted components with higher indices are also truncated, which leads to effective denoising of the data.

A specific area of interest in inelastic spectroscopy-based modalities is the ability of MV techniques to spectrally \emph{unmix} the data. Taking the example of a Raman dataset, it is often observed that the spectrum collected from a single spatial pixel contains information of a mixture of different components inside the (biological) material being probed \cite{Prats-Mateu2018}. In Brillouin imaging, the typical transverse resolution is $1~\mu m$, and the axial resolution tends to be much worse \cite{Fiore2019a}. These dimensions are large compared to some of the features in biological material and produce scenarios where, a single focal volume was influenced by more than one phase in the sample \cite{Mattana2018}. This is equivalently stating that the focal volume tested is not \emph{pure}, thus could be unmixed into composite structures of different spectral signatures, i.e. different endmembers. The computation then performed by any unmixing MV analysis will be to estimate the abundance of each endmember at every \emph{voxel}, in a supervised fashion, if the total number of endmembers is known or otherwise unsupervised, when this has to be estimated statistically \cite{Ceamanos2016}. PCA thus lends itself naturally as a statistical approach to spectral unmixing. In practice, the number of components $p$ is a key parameter that is usually determined by prior knowledge of the sample. Since the decomposition procedure is independent of $p$, a purely unsupervised approach is possible by thresholding the percentage variance that is covered by a certain number of components, although it is difficult to set a universal threshold that is applicable to all samples. Conversely, the same property also means that PCA is inherently ignorant of the true underlying endmembers. For this reason, albeit easily accessible, the most basic form of PCA does not offer the best solution for spectral unmixing in Brillouin data, as it is sensitive to common spurious features, such as fluctuation in the elastic peak or laser drift, and thus requires data pre-processing.

\subsection*{vertex component analysis}
A more robust approach of component analysis, especially for spectral unmixing, comes in the form of vertex component analysis (VCA). Compared to the statistical approach of PCA, this approach is geometrical and relies on the assumption that the HS data can be viewed to span a \emph{simplex} in an n-dimensional Euclidean space, where each frequency shift value from the spectra can be plotted. A simplex can be a generalised triangle or tetrahedral geometrical shape for example, containing $n+1$ vertices in an n-dimensional space. Additional assumptions will then have to be made to identify the endmembers in the simplex and may vary between algorithms. Nonetheless, the one assumption that is always made in this analysis is that the (spectra of) endmembers are located at the vertices. In terms of spectral unmixing, there are two main approaches, both aiming to solve an optimisation problem. The first kind assumes that there is always at least one pure voxel per endmember, which suggests it is always possible to find a position in the Brillouin image that only contains the spectral information of one species, for every species present. The mathematical treatment of this kind of algorithms is rather complex and the reader is referred to Ref. \cite{Nascimento2003} for a thorough treatise. Qualitatively, a simplex is fitted to the dataset first to span the maximum volume, whose vertices correspond to a combination of the purest voxels \cite{Bioucas-Dias2012}. If, however, every image pixel is non-pure, which may well be the case for complex biological structures encountered in BI, it is then more appropriate to change the initial assumption and fit a simplex with the minimum volume \cite{Li2008}. This approach retrieves the endmember spectra perfectly in the absence of noise, even if they cannot be observed in the actual data \cite{Li2015}. In both cases, the mathematical representation of the analysis is that:
\begin{align}
\hat{\mathbf{D}}&=\mathbf{AM^T}+\hat{\mathbf{N}}\\
\mathcal{C}&=conv\lbrace M \rbrace
\end{align}
where the $\mathbf{M}\in \mathbb{R}^{N_s\times p}$ is now the mixing matrix that contains the spectra of $p$ endmembers and similar to PCA, the abundance of each is estimated by the normalised matrix $\mathbf{A}\in \mathbb{R}^{N_xN_y\times p}$, which is equivalent to the PCA scores and can be used to form abundance images. The simplex fitting of $p$ vertices is carried out over $\mathcal{C}$, which is the convex hull enclosing the columns of $M$ in spectral space. An additive noise model is applied again as the residual variance is explained by the noise matrix $\hat{\mathbf{N}}$. The VCA algorithm used in this work is open-source and available as a \emph{Matlab} function provided from Ref.\cite{Li2015}, though minor modification was made to implement it with the Biodata protocol. For reasons discussed above, VCA is more advantageous than PCA for the purpose of unmixing, particularly in situations where the spectral data consist of quite complex features (i.e. numerous endmembers) and generally exhibit superior performance both in terms of computational speed and accuracy. While the majority of the process is unsupervised, for Brillouin data the most important factor for successful unmixing relies on the choice of $p$. Unlike for PCA, the choice of $p$ plays an important role in the computation here, as the shape of the endmember spectra retrieved directly depends on the number of vertices fitted. Finally, it is noted that the aforementioned pre-processing techniques can still be applied to the data before VCA when appropriate and does not change the outcome.

\subsection*{linear discriminant analysis}
So far, component-based multivariate methods have been introduced with a focus on their ability to extract spectral features in a sub-pixel mixture of different species. It is also possible to make the opposite assumption and \emph{classify} each spectrum in the dataset to be strictly due to one species only. To achieve spectral classification, i.e. the labelling of each spatial pixel in the image, it is mandatory to have some \emph{a priori} knowledge of the sample under-test. In this vein, linear discriminant analysis (LDA) stands out as a common supervised method with high performance. Numerous analogies can be drawn between LDA and PCA, although as the name suggests discriminant analysis concentrates on modelling the data into any given number of classes in order to generate the maximum contrast between classes, rather than those that best describe the data. Mathematically, in LDA the grouping of data is realised by defining two metrics, the intra-class variance ($S_{I}$) and the inter-class variance ($S_{II}$):
\begin{align}
S_I&=\sum_{j=1}^{c}\sum_{i=1}^{N_j}(n^j_i-\mu_j)(n^j_i-\mu_j)^T\\
S_{II}&=\sum_{j=1}^{c}(\mu_j-\mu)(\mu_j-\mu)^T
\end{align}
where for $c$ total classes and $N_j$ total samples for the jth class, $S_I$ is constructed by calculating the difference between $n^j_i$, the ith sample of the jth class, and $\mu_j$ the mean of the jth class. Similarly, for the inter-class matrix, $\mu_j$ is compared to $\mu$ the overall mean of all classes. The objective of the numerical implementation is to maximise the inter-class metric, as well as minimising the intra-class metric in tandem, usually by solving the maximisation problem: 
\begin{equation}
\mbox{max}\big\{ \dfrac{det(S_{II})}{det(S_{I})} ~|~det(S_I)\in \mathbb{R}_+^n\big\}.
\end{equation}
An immediately foreseeable problem arises when the intra-class variance is a singular matrix, which means $det(S_I)=0$. Unfortunately, spectral matrices are almost always singular \cite{Hedegaard2010}. As such, PCA and VCA serve an additional purpose as both can transform the raw data into an intermediate space, described by a hyperplane where each image pixel can be represented by the scores of chosen vertex or principal components. In this intermediate space, it is then possible to define $c-1$ discriminant functions to sort the data into $c$ classes \cite{FISHER1938}. Other algorithms, such as partial-least-squares (PLS) regression-based discriminant analysis \cite{Barker2003}, avoid the singularity problem by performing dimension reduction and classification in a single step, although their performance is not necessarily superior \cite{Rezzi2005}. %As mentioned before, classification is only possible with the input of a labelled dataset as a training set and the model built upon this data can be used to classify future data with the decision boundaries computed, which are vectors orthogonal to the hyperplane that satisfy the maximisation criterion. While this boundary is frequently defined by a linear function, giving the method its name, it is also possible to define other functional shapes, such as in quadratic discriminant analysis (QDA) \cite{Srivastava2007}.
To perform LDA on our data, we combined functions from the \emph{Statistics and Machine Learning Toolbox} in \emph{Matlab} with Biodata. Even though supervised approaches like LDA are 'tried-and-true' methods for classification, they rely strongly on the amount and quality of the training data \cite{Martinez2001}, which is a challenging aspect for Brillouin imaging due to the scarcity of data in general. Additionally, the real potential of this technique extends beyond just identification of species within a single image, as with good training it can be used to diagnose diseased organisms with high sensitivity and specificity, which is already routinely practised in data generated by vibrational spectroscopies \cite{BakkerSchut2000,Kendall2003,Krafft2007}.

\subsection*{hierarchial cluster analysis}
Instead of labelling the data with pre-knowledge of the sample, there are completely unsupervised algorithms that can independently generate these labels. Cluster analysis is the prime example of such techniques and a common choice in the analysis of hyperspectral data in other applications. In HS data, the most basic cluster analysis will first need to calculate the pairwise distance between every possible pair in the dataset, which is the most computationally intense part of the algorithm. For example, for a dataset with 1000 spectra, this would lead to a total of $^{1000}C_2=499500$ distances being computed. How the distance metric is defined is clearly of importance for the success of meaningful clustering and will be discussed in more detail below. Qualitatively, the next step in the process then starts to link observation pairs in the data that are closer together into clusters. Once all the data are in clusters, the closest clusters will then be joined to form even larger clusters, this process is commonly repeated algorithmically using Ward's method \cite{Ward1963} in a `bottom-up' fashion until one mega-cluster is formed at the top. Identification of natural clusters in this fashion is known as agglomerative hierarchical cluster analysis (HCA).
%As an illustration, a real dendrogram generated in analysing the same Adipocyte data from before is plotted below, where the colour-coding correspond to a scenario where the threshold has been set to group the data into 6 distinct clusters.
%\begin{figure}[h!]
%	\centering
%	\includegraphics[width=\textwidth]{FIGURES/ch5/dendrogram}
%	\caption{A dendrogram showing the clustering of real Brilloun HS data collected from an Adipocyte. The 6 clusters formed are colour-coded accordingly.}
%	\label{fig:dendro}
%\end{figure}
In the context of analysing Brillouin data, there are practical considerations when applying HCA. For instance, the total number of clusters formed $C$ is a free initial parameter and controls the so-called granularity of the clusters, and should not be confused with the total number of endmembers expected in terms of spectral unmixing. Since for cluster analysis pixels are always pure, it was found that for Brillouin data $C$ should be set high enough initially to retain all spatial features in the data, for consistency, $C$ is thus fixed at 8 for all cluster analyses performed in this work. Sub-grouping of different clusters can then be performed by looking at the mean spectra of the initial clusters, which enables a second round of clustering that yields the total number of endmembers expected from the particular dataset. Ultimately, a hyperspectral image can be formed by assigning different \emph{segments} of the image to different clusters, which can be colour-coded for high contrast. This absolute partition of data may become challenging in samples which are more complex i.e. spectrally mixed, and thus clustering methods that do not practice this clear-cut spectral segmentation do exist, such as Fuzzy C-means clustering \cite{Dunn1973}. As it was inferred earlier, the other important factor for successful clustering is the definition of the distance metric, which is a measure of similarity between spectra. This pairwise distance can be calculated in a variety of ways, each with its advantages and disadvantages. The most straightforward distance measure can be the most literal definition, which is given by the physical distance between the spectra pair in Euclidean space. For a HS dataset with the dimension ($N_xN_y\times Ns$), the distance $d_{ij}$ between two spectra $\mathbf{S_i}$ and $\mathbf{S_j}$, each spanning $Ns$ data points is given by the generalised form:
\begin{equation}
d^{(R)}_{ij}=\left(\sum_{m=1}^{Ns}|\mathbf{S_{im}}-\mathbf{S_{jm}}|^R\right)^{1/R}
\end{equation}
This is known as the Minkowski distance and it should be realised that the standard Euclidean distance simply appears as a special case for when $R=2$, which is the most common and effective measure used in our HCA algorithm. In such normed Euclidean spaces, there are multiple other distance measures that may be defined in terms of the Minkowski distance, for example the city-block distance ($R=1$) and Chebychev distance ($R=\infty$) \cite{Rokach2010}. 

For a more statistical definition, the distance can also be measured in terms of the linear Pearson correlation coefficient \cite{statsbook2} or the Mahalanobis distance between spectra, respectively:
\begin{equation}
d^{(C)}_{ij}=1-\dfrac{(\mathbf{S_i}-\bar{\mathbf{S_i}})-(\mathbf{S_j}-\bar{\mathbf{S_j}})}{\sqrt{(\mathbf{S_i}-\bar{\mathbf{S_i}})\cdot(\mathbf{S_i}-\bar{\mathbf{S_i}})^T} \sqrt{(\mathbf{S_j}-\bar{\mathbf{S_j}})\cdot(\mathbf{S_j}-\bar{\mathbf{S_j}})^T}}
\end{equation}
\begin{equation}
d^{(M)}_{ij}=\sqrt{(\mathbf{S_i}-\mathbf{S_j})\cdot\hat{\mathbf{C}}^{-1}(\mathbf{S_i}-\mathbf{S_j})^T}
\end{equation}
where the $~\bar{}~$ notation is used for mean values, $\hat{\mathbf{C}}$ is the covariance matrix and all other symbols retain the same meaning. Pairwise correlation is the most physically intuitive measure in the ideal case, as it gives a quantitative measure of similarity between spectra. In a low SNR regime, however, the extraction of the pairwise correlations is not an easy task and is rigorously covered elsewhere \cite{Moriya2008}. Mahalanobis distance, on the other hand addresses the issue of noise. It measures distance between a pair of observations by quantifying their separation in terms of number of standard of deviations \cite{Mahalanobis1936}, which is essentially a consistency test with some noise tolerance. To implement HCA, the \emph{Statistics and Machine Learning Toolbox} also provides relevant functions which allow the calculation of all the distances discussed above and their linkage using Ward's algorithm. After integration with the Biodata protocol, performing HCA on our datasets was noticeably the slowest of all multivariate techniques and the computation time scales with the number of clusters $C$ being searched for. Parallelisation \cite{Li1998} can be implemented to further improve the analysis speed in the future.

\subsection*{k-means cluster analysis}
A direct improvement on the computational efficiency of HCA comes in the variant form of k-means cluster analysis (KCA). The core concepts of KCA are identical to those of HCA, as different distance measures can still be used to group spectra into clusters. The main difference, is that KCA treats the clustering problem using a centroid-based approach, which requires the random selection of $K$ spectra from the data as starting positions. By doing so, only $N_s\times K$ initial distances need to be computed, thus leading to faster clustering. These spectra become centroids of the potential clusters and the remaining spectra can each be affiliated with one of these centroids according to the distance measure chosen. Once all spectra have been assigned to clusters, the centroids of the $K$ clusters are re-calculated and the assignment process is repeated in an iterative manner, with the termination condition being threshold on the amount of changes in the assignment per iteration or simply on the number of iterations. To enable direct comparison, the number of starting centroids $K$ is also fixed at 8 for all KCA carried out, although multiple attempts may be needed to achieve the best results due to the occasional clumping of multiple clusters, caused by closely-spaced initial centroids which are chosen randomly. This is also implemented in \emph{Matlab} using the same toolboxes as before, and the KCA algorithm in this scenario is simply a more efficient form of HCA, albeit losing some freedom in the choice of similarity measures, which can be custom-defined and incorporated for future work if needed. 
%\subsubsection*{density-based cluster analysis}

\subsection*{cluster analysis with multi-modal data}
A major limitation in the cluster analysis of Brillouin data is the existence of spurious features and noise in the data. In view of the degradation that the final cluster images suffer from, some optimisation strategies have also been tested. To specifically address the noise issue, Ester et al. \cite{EsterM1996} have devised a density-based method which should achieve cleaner partitioning even in the presence of noise. This technique demonstrated qualitative improvement to our data, the results can be found in Supplementary materials. For further enhancement, we propose a new approach in which the inclusion of \emph{a priori} information on the sample can be used to train the HCA algorithm.

A common practice in existing BI techniques is to accompany the hyperspectral images of the sample with a white light image as reference, which in most cases can be co-registered. This image contains \emph{a priori} information, specifically intensity-based information that can be fed to the HCA algorithm. To incorporate this additional knowledge, a new algorithm was designed in \emph{Matlab}, whose flow can be found below in Figure \ref{fig:clusterflow}. 
\begin{figure*}[h!]
	\centering
	\includegraphics[width=\textwidth]{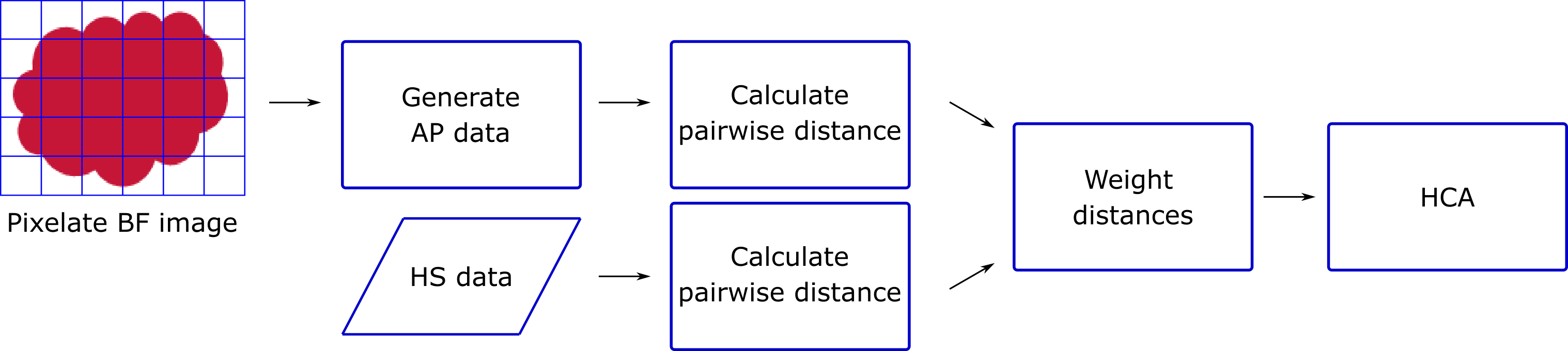}
	\caption{A flow chart depicting the algorithm designed to include AP information for cluster analysis of hyperspectral data.}
	\label{fig:clusterflow}
\end{figure*} 
This simple approach first selects a region-of-interest (ROI) from the bright field image and downsamples the image by software into the dimension that matches the hyperspectral image, which is $N_x\times N_y$. This leads to a lower resolution image, but most of the image contrast should be retained, this image will be named the \emph{a priori} Image (API) from here on. The API is then one-dimensionalised into a vector of dimension $N_xN_y\times 1$. This way a fictitious `hyperspectral' dataset is generated, whereby each `spectrum' contains intensity information from the API. The next step calculates the pairwise distances, according to a chosen metric, for both the AP dataset just generated and the actual HS dataset obtained from experiment. Since the hyperspectral image should be structurally similar to the API, the distances should resemble one another and large differences can be assumed to be due to noise. Thus, the effect of noise can be mitigated by utilising a weighted sum $w$ of the two distances. If the normalised distance vectors from the AP and experimental data are respectively denoted by $\mathbf{d_{AP}}$ and $\mathbf{d_{HS}}$, the weighted distance measure $\mathbf{d_{w}}$ will be given by:
\begin{equation}
\mathbf{d_{w}}=(1-w)\cdot\mathbf{d_{HS}}+w\cdot\mathbf{d_{AP}}
\end{equation}
Finally, clusters can be formed by pairing the distances in $\mathbf{d_{w}}$ by using standard HCA. To investigate the effect of this intensity model-based algorithm, the 2nd order Minkowski distance was chosen with a weighting factor of $w=0.2$ on data collected on single cells (adipocytes).

\section{Methods and materials} \label{sec:exp}
\subsection{sample preparation}
The phantom used in this work was a blend of hydrogel and polystyrene beads, which were chosen due to their highly contrasting mechanical properties. The hydrogel was prepared following a reported protocol (Ref. \cite{Tse2010}). Polystyrene beads with a specified outer diameter of $\sim90\mu m$ were embedded in the gel during preparation. The preparation procedure ensured that there were various fixed beads at different depths of the phantom.

The live cells were cultured and provided by Dr. Cristoforo Silvestri and according to her previously published protocol \cite{Silvestri2013}. Adipocytes cells are also known as fatty cells, whose main function is to store energy as fat \cite{Birbrair2013}. Thus this cell mostly consist of lipids droplets which is a main constituent of fat, as well as any other common cellular structures, such as cytoplasm, cell membrane and nucleus.

\subsection{confocal Brillouin imaging}
To perform Brillouin imaging on the phantom and cells, a custom-built confocal Brillouin microscope was used, whose technical details have  already been reported elsewhere \cite{Karampatzakis2017}. The implementation of an interferometric filter \cite{BRI_Carl2016} was needed to suppress stray light in the spectrometer for sufficient contrast, a laser source at $561nm$ (\emph{Cobolt Jive}) was utilised in conjunction with a 20X objective (\emph{OLYMNPUS PLN 20X Objective}, NA=0.4) for the phantom and a 60x objective (\emph{OLYMNPUS UPLFLN 60X Objective}, NA=0.9) for cells, which are expected to produce a diffraction-limited spot size of $1.3\mu m$ and $0.6\mu m$ respectively. For the phantom, the field of view (FOV) was selected to include just one bead and first imaged using bright field microscopy for reference, the same FOV was then probed using the Brillouin microscope set-up with a sampling step size of $5\mu m\times5\mu m$ to co-register and produce a HS dataset of dimension $30\times30\times200$. For adipocytes, one particular cell in the bright field FOV was chosen for each experiment, the ROI was defined by a $45\mu m\times50\mu m$ window which was raster scanned with a pixel size of $1\mu m\times1\mu m$. Due to large thermal fluctuations in the laboratory caused by mitigating circumstances, the filter had to be reset at regular intervals to account for the thermal drift in the laser. As a result, each cell was imaged in 8 stripes, which were then stitched together using software during post-processing.

\section{Results}\label{sec:results}
\subsection{phantom}
\begin{figure}[h!]
	\centering
	\includegraphics[width=\textwidth]{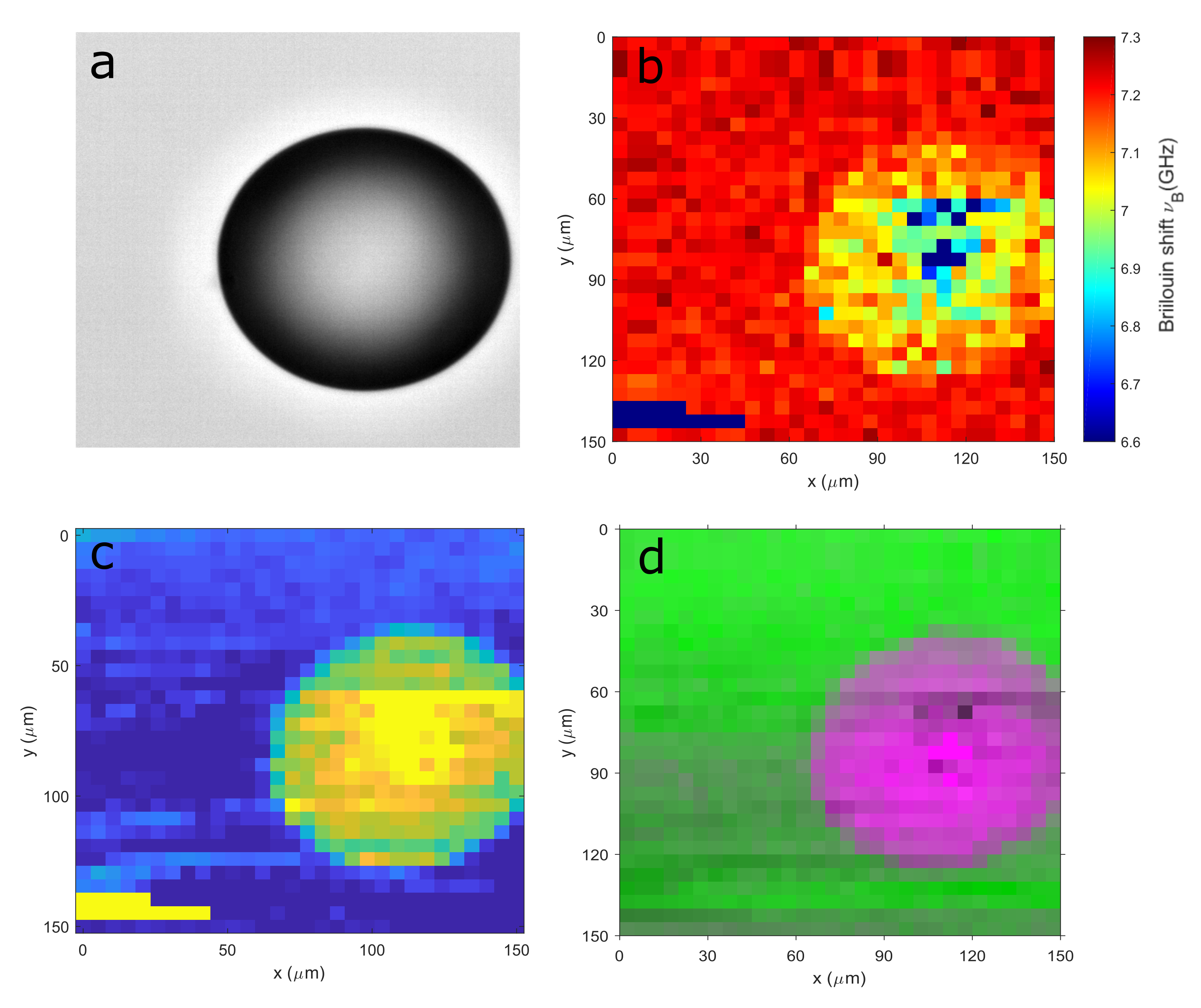}
	\caption{\textbf{Images of the phantom.} \textbf{a} A bright field image indicating the FOV for BI. \textbf{b} Brillouin image obtained by mapping the fitted Brillouin shift values. \textbf{c} Abundance image formed by using PCA, mapping the scores of PC3. \textbf{d} High-contrast image formed with multiple components obtained from VCA, cyan is used for the polystyrene endmember and green for hydrogel.}
	\label{fig:sphere1}
\end{figure}
\noindent To examine the effect of the multivariate algorithms in practice, a ROI in the phantom (Figure 2a) was imaged for data analysis. In order to enable a direct comparison, the conventional routine of Lorentzian fitting was also utilised to extract the peak shift and width of each spectrum in the dataset. Background subtraction of the Rayleigh peaks was first performed during pre-processing. As the Stokes and Anti-Stokes Brillouin peaks are symmetrical and should contain the same information, two Lorentzian profiles between the free spectral range (FSR) of $0-30$ GHz were fitted to calculate an average frequency shift. The fitting was deemed good enough according to a $R^2$ statistical test ($\ge0.95$), the Brillouin shift/width values of all 900 data points were retrieved in $\sim 120s$ with a standard workstation (\emph{Intel(R) Core(TM) i7-3770 @ 3.40 GHz 3.90 GHz, 16 GB RAM}). From the Brillouin shift heat map presented in Figure 2b, the first observation to be made is that some contrast was obtained by direct line fitting, as the general shape of the spherical bead was evident, and simple visual estimation would validate the expected dimension of $90\mu m$. In terms of the mechanics, using the link of the Brillouin interaction to the bulk compressibility of the sample \cite{Lai2010}, the polystyrene beads are expected to produce a significantly higher frequency shift value in comparison to the surrounding hydrogel. On the contrary, a lower shift was observed in the region enclosed by the plastic sphere, seemingly suggesting that the bead is actually more compressible. Another feature that is likely an artefact came in the shape of some abnormally low-shift pixels at the corner and centre of the sphere, which might point towards defects in the phantom. These hypotheses cannot be ruled out without further spectral analysis and MV algorithms provides the means to quickly study the relevant spectra.

As a direct comparison, the same data was then analysed using PCA, initially assigned three principal components (PC), to account for the two main mechanical indicators (hydrogel and polystyrene) and their interface. The decomposition of the entire dataset only took $600ms$ with the same workstation from before and could already be re-arranged into images of abundance for individual endmembers. By looking at the spectral profiles of these loadings, it was apparent that the first component, which accounts for the largest variation in the data was caused by the Rayleigh background. As explained previously, this result is expected from PCA algorithms as they are intrinsically unable to tell background fluctuations from otherwise useful spectral differences. The second and third components, however, resembled Brillouin spectra of hydrogel and polystyrene and can be found in the Supplementary materials. PC2 was fitted to retrieve an average shift value of $7.19\pm0.1$, which coincides with reported hydrogel values of this particular variant and wavelength \cite{BRI_kathy2018}. Interestingly, PC3 contained two sets of peaks, the first pair of peaks were weaker in intensity and produced an average shift of $7.16\pm0.1$ GHz. The other set of higher-shift peaks produced an average shift of $13.3\pm0.2$ GHz, which is closer to what is expected from polystyrene. This also explains the low-shift values previously observed inside of the sphere. The fitting algorithm used requires the definition of certain initial parameter to function optimally, including the number of peaks to be fitted. In this case, the residual hydrogel peaks were selected by the algorithm instead of the more representative higher frequency pair and the slight decrease in the mean shift can be attributed to the subtraction of the Rayleigh background at low SNR, which may cause the slight distortion in the peaks at round $\pm7$ GHz. In effect this observation echoes an earlier argument for the need of spectral unmixing. Specifically, since the Brillouin signal was collected using a relatively low magnification objective ($NA=0.4$), the collection volume likely included back-scattered photons from both hydrogel and polystyrene during acquisition. Close inspection of the reference bright field image also indicates strong reflection, which suggests that the focal plane was near the top surface of the bead and the focal volume may include some hydrogel content. As further validation, an abundance map was reconstructed using only PC3 of polystyrene (Figure 2c), where comparatively better contrast is evident. The spherical shape of the bead is now clearly defined as the interface region is now visible due to the spectral features there resembling both the hydrogel and bead component, thus gaining an intermediate abundance. On the other hand, the spurious features at the centre of the bead and the corner of the ROI are still visible.

For improvement in spectral unmixing, VCA was carried out by decomposing the data further using four vertex components (VC), the computation time for which also further decreased to $300ms$. As a result, the additional component managed to separate the spurious artefacts and examination of the endmember spectrum clarified that these were caused by stray light, which saturated the detector. This was most likely due to the strong reflection at these spatial positions, which were not fully suppressed with the interferometric filter, nor the software filter during pre-processing. Since this is a known challenge associated with the instrumentation, no further discussion will be given here with regards to treating stray light. Having separated the stray light component, a high contrast image can be formed by combining different components. The result can be found in Figure 2d, where the abundance of first three components were combined. The spectra at the vertices can also be used as signature profiles for the endmembers, and it was found that two components were spectrally very similar, only an overall offset in the Brillouin peak positions was measured. The abundance maps of these two components effectively divides the image into two halves, with each component having higher abundance in the background regions of the bottom and top half respectively. Since the data was obtained by raster scanning from the bottom, both components were attributed to be characteristic of hydrogel and the difference was explained by the frequency drift of the laser, which caused a linear shift in the spectra. By averaging these two components to produce a single spectrum, it was only necessary to fit a total of just two endmember spectra to obtain mechanical information on hydrogel and polystyrene, the errors on the parameter estimation were calculated via bootstrapping \cite{Singh2010}, the results are summarised in Table 1.
\begin{table}[H]
	\begin{center}
		{
			\begin{tabular}{|l||c|c|}
				\hline
				& Brillouin shift $\nu_B$ (GHz) & linewidth $\Gamma_B$ (GHz)\\ \hline
				Hydrogel &  $7.194\pm0.036$ & $1.31\pm0.20$ \\ \hline
				Polystyrene & $13.30\pm0.026$    & $0.92\pm0.09$  \\ \hline
			\end{tabular}
			\caption{Brillouin information of the two identified endmembers in the spectral mixture, obtained by Lorentzian fitting and bootstrapping. \label{tab:sphere} }
		}
	\end{center}
\end{table}
\noindent Additionally, spectral classification and segmentation were also performed on the same dataset. For a supervised study, typical LDA can be implemented by using a holdout approach. Specifically, 450 out of the total 900 spectra, which correspond to the bottom half of the image, were used to train a linear model. To reduce the data dimension, the scores of the VC that corresponds to polystyrene was chosen as the input in the intermediate component space, where initial labelling was completed by dividing the data into two groups by the mean value. Clear outliers ($\ge2$ standard deviations away from the mean) were first removed and those remained possessing a score that is higher than the mean would be labelled `polystyrene' and vice versa by 'hydrogel'. By applying this model on the unseen, second half of the data, the remaining points were classified with a binary decision, the resultant image of the sphere reconstructed from the classification can also be found in the same figure. Analogously, the entire data cube can also be segmented in an unsupervised fashion with HCA. In view of the large spectral differences between hydrogel and polystyrene, a simple city-block distance was used to form just 3 clusters in about 1 second. The final image is compared alongside that obtained from LDA, and it can be seen that the outliers (in green) were not removed by unsupervised learning, which highlights the dependence of the choice of algorithm on the data quality.
\begin{figure}[H]
	\centering
	\includegraphics[width=\textwidth]{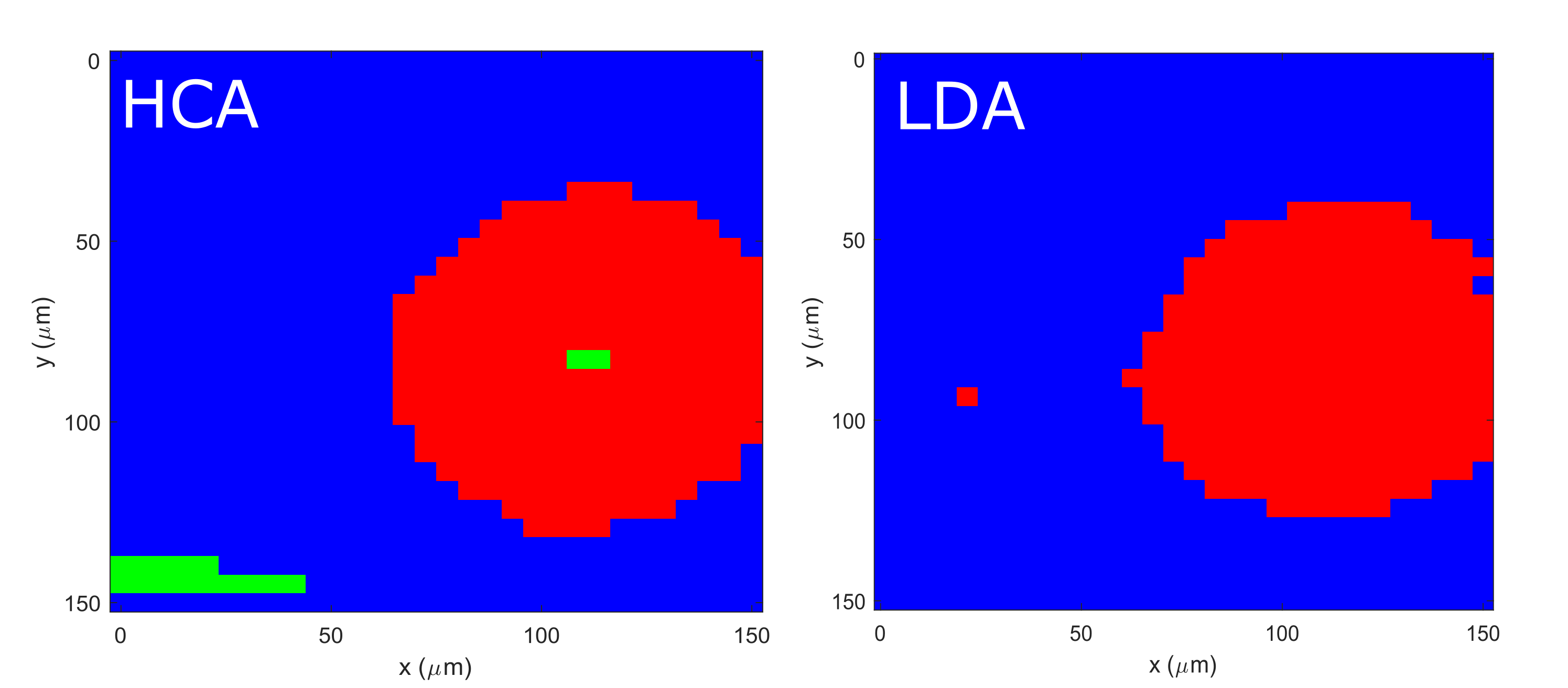}
	\caption{\textbf{Spectral labelling using supervised and unsupervised learning.} \textbf{(LEFT)} RGB image of the phantom obtained using supervised LDA. \textbf{(RIGHT)} RGB image of the phantom formed using unsupervised HCA. Red is used indicate the class for polystyrene, blue for hydrogel and green for outliers.}
	\label{fig:sphere2}
\end{figure}

\subsection{3T3L1 adipocyte}
\begin{figure}[h!]
	\centering
	\includegraphics[width=\textwidth]{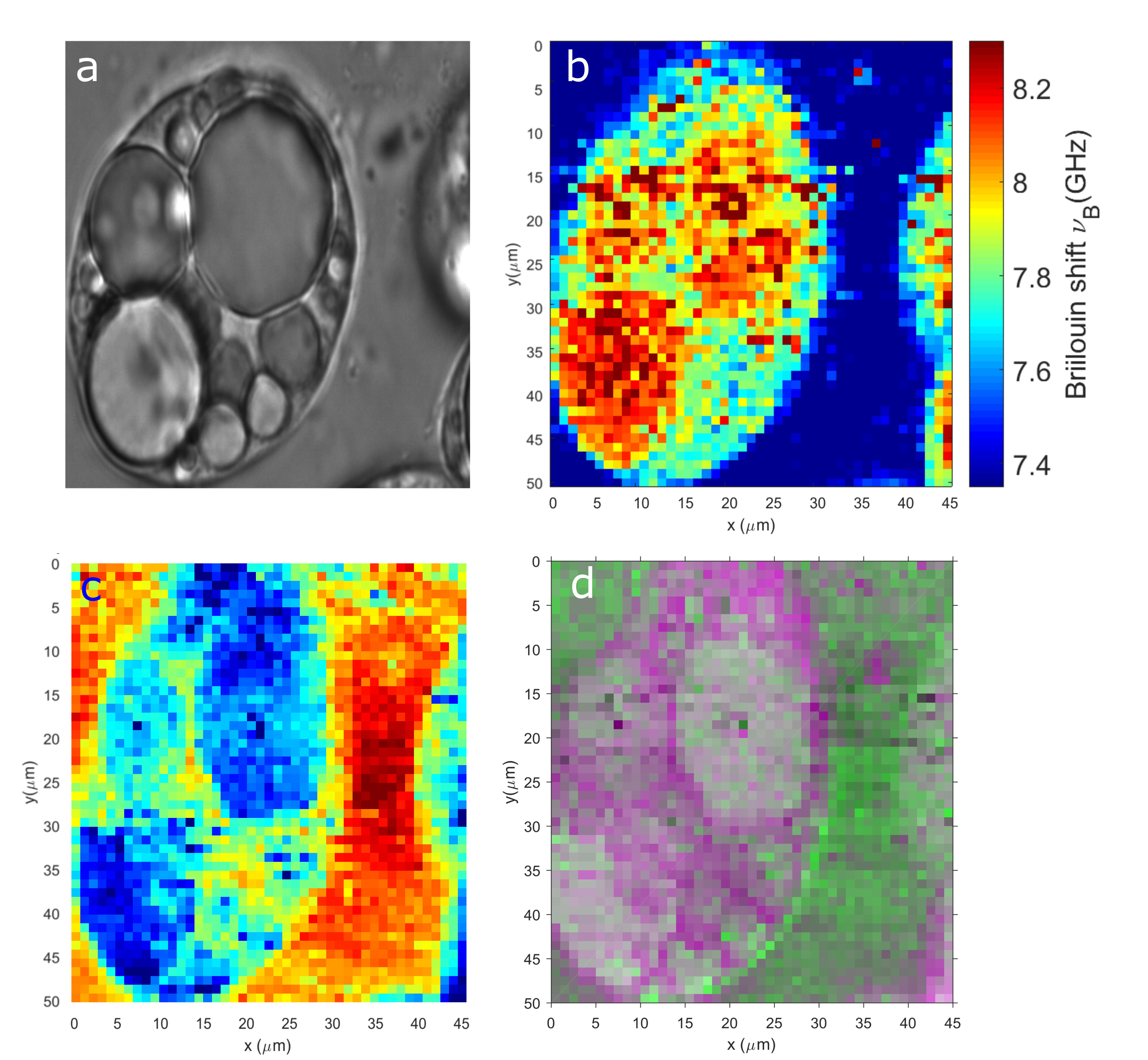}
	\caption{\textbf{Images of a 3T3L1 fat cell}. \textbf{a} A reference image for BI obtained using wide-field bright light microscopy. \textbf{b} Brillouin image reconstructed by mapping the average shift each pixel, obtained by Lorentzian fitting. \textbf{c} Abundance image for VC3, the unmixed endmember that corresponds to PBS. \textbf{d} Composite image consisting of the three identified spectral components VC2, VC3, VC4, the difference image of VC2 and VC3 was taken first and then combined with VC4 to form the final image using a bi-colour scheme.}
	\label{fig:cell1}
\end{figure}
Similar to before, images of single adipocytes were first formed using traditional fitting and then directly compared to those obtained from MV analysis. Due to the need to combine data strips, additional measures were called for in order to remove the stitching artefacts in the images. When using the intensity fitting method, this was easily remedied by fitting both the Stokes and Anti-Stokes peaks for the same order, whereby an average can be calculated and hence removing any linear effect due to the laser drift. Equivalently, the Brillouin peak of the immersion liquid, in this case Phosphate-buffered saline (PBS), could be used as a reference to correct for the drift, similar to the technique used in \cite{Bevilacqua2018}. For the component-based analysis, the abundance for each component was first obtained for each strip and normalised using the SNV transform to correct for any relative drift. The normalisation does, however, slightly decrease the contrast of image. The Brillouin image obtained by least-squares fitting is presented in Figure 4b, the frequency shift values of all $2250$ pixels were retrieved in around 5 minutes with the same desktop as before, with certain points producing less trustworthy values statistically ($R^2<0.8$) and was thus replaced with the average value of the last two pixels. It is quite clear from this image that an overall contrast was again achievable by simply displaying information of a single peak. The background liquid is well separated from the cell content and furthermore, the lipids generally exhibit a higher shift value, which give rise to the sharp contrast in the image. The image does appear noisy, however, and not all features from the original white light image (Figure 4a) were retained in this Brillouin equivalent.

VCA with four vertices was performed on the same data in around $800ms$, the additional normalisation process did not significantly increase this processing time. The resulting endmember spectra are presented in the Supplementary materials. The first endmember was still attributed to the shifting of the Rayleigh orders, and was thus removed from the data as the background. To facilitate the discussion of each remaining spectral component, their average shift and linewidth values obtained from Lorentzian fitting are summarised below in Table 2.
\begin{table}[h!]
	\begin{center}
		{
			\begin{tabular}{|l||c|c|}
				\hline
				& Brillouin shift $\nu_B$ (GHz) & linewidth $\Gamma_B$ (GHz)\\ \hline
				VC2 &  $8.02\pm0.18$ & $2.64\pm0.54$ \\ \hline
				VC3 & $6.79\pm0.22$    & $0.59\pm0.43$  \\ \hline
				VC4 & $7.57\pm0.20$    & $1.10\pm0.50$  \\ \hline
			\end{tabular}
			\caption{Estimated Brillouin parameters for the obtained vertex components. \label{tab:cell1} }
		}
	\end{center}
\end{table}
These values are consistent with those calculated from conventional fitting for lipids, PBS and cytoplasm \cite{wu2020}. It is noted, however, that the fitting errors are high, especially for the linewidth estimates due to the lack of any constraint for endmember profiles to stay Lorentzian, which may be an area of algorithmic improvement. To verify these findings, images can be formed from the individual components to check for `hotspots', where these mean shift values are most applicable. As an example, the hypothesised PBS component was used to reconstruct an abundance image in Figure 4c. The background region displayed the highest abundance of this particular component, which is indeed where PBS is expected to be. Meanwhile, an enhanced contrast is observed as the abundance decreases moving into the cell, and virtually drops to zero where the lipid droplets are expected. Notably, smaller droplets fo lipids that were visible in the original BF image have been recovered. The same process was repeated for the remaining components and it was concluded that they did indeed correspond to the region of cytoplasm and lipids. Combining the VCA results, just as it was previously done for the phantom, a composite image was reconstructed and shown in Figure 4d to highlight previously hidden details. 

In terms of unsupervised learning, cluster analysis was performed on a different adipocyte to extract underlying patterns. Using two rounds of spectral matching as described previously, a segmented image of the cell was obtained in $3.90$s in the case of HCA and in $0.83$s when using KCA. Two different distance measures were used to illustrate the effects of the similarity metric on image segmentation, namely the second order Minkowski distance and the Mahalanobis distance. Resultant RGB images are presented in Figure 5b and 5c respectively. For quantitative analysis, the average spectra of each segment (see Supplementary materials) of the image were fitted for both cases, the fitting results can be found in Table 3.
\begin{figure}[h!]
	\centering
	\includegraphics[width=\textwidth]{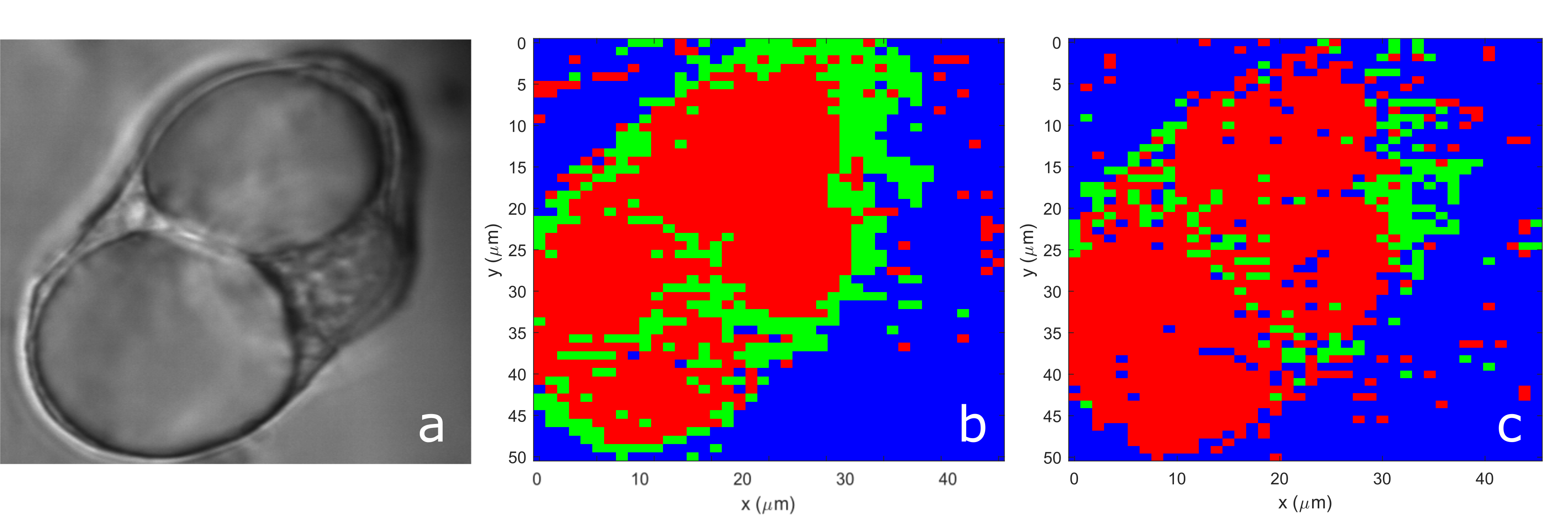}
	\caption{\textbf{Spectral segmentation of a single adipocyte with cluster analysis}. \textbf{a} Reference image of the cell studied. \textbf{b} RGB image of the cell, obtained by HCA using the order 2 Minkowski distance. \textbf{c} RGB image of the cell, obtained by HCA using the Mahalanobis distance. Red, green and blue denote the regions for lipids, cytoplasm and PBS respectively.}
	\label{fig:cell2}
\end{figure}
\begin{table}[h!]
	\begin{center}
		{
			\begin{tabular}{|p{3cm}||c|c|}
				\hline
				Minkowski (R=2) &  $\nu_B$ (GHz) & $\Gamma_B$ (GHz) \\ \hline
				PBS &  $7.19\pm0.008$ & $1.34\pm0.02$ \\ \hline
				Cytoplasm & $7.65\pm0.006$    & $1.83\pm0.02$  \\ \hline
				Lipids & $7.90\pm0.036$    & $2.58\pm0.18$  \\ \hline
			\end{tabular}
			\begin{tabular}{|p{3cm}||c|c|}
				\hline
				Mahalanobis &  $\nu_B$ (GHz) & $\Gamma_B$ (GHz) \\ \hline
				PBS &  $7.27\pm0.033$ & $1.42\pm0.02$ \\ \hline
				Cytoplasm & $7.65\pm0.018$    & $2.13\pm0.10$  \\ \hline
				Lipids & $7.80\pm0.04$    & $2.62\pm0.19$  \\ \hline
			\end{tabular}
			\caption{Estimated Brillouin parameters for the clusters as identified by the two different distance metrics. \label{tab:cell2} }
		}
	\end{center}
\end{table}
There are discrepancies in the values obtained via the two measures which cannot be explained by their respective errors, and this is especially true for the retrieved linewidth values. It can be seen from the images that both measures were able distinguish the main contrast in the cell through the spectral differences between the PBS, cytoplasm and lipids. The version that was created using the Euclidean distance was arguably the better out of the two as it appeared less prone to background and noise, albeit still suffering from stitching artefacts. In contrast, the use of Mahalanobis measure managed to capture the key features of the cell without artefacts, although the features were likewise poorly defined due to its incapacity to differentiate small changes in the spectra which are comparable to the standard deviation threshold. This may explain the fuzziness in the image, and may be improvable by tuning the constraint for clustering or denoising first. While the two cells are not required to have identical spectral parameters, the Brillouin properties of PBS is expected to stay constant in the sample. It is observed that with the errors considered, there are also small differences between values obtained from HCA and previously from VCA for PBS. In part, it is because cluster analysis produces a simple statistical average of existing data for each cluster, thus retains the spectral shape of the data, which is potentially a relative advantage when considering the fidelity of linewidth analysis. On the other hand, the higher shift values are due to the misclassification of some pixels that are visible in the images, which appear in the lipid regions and hence explains the increase in the mean values calculated for PBS. Fundamentally, it also reflects on the differences of clustering methods, due to their pure pixel assumptions, which indicates that one limitation of using HCA in Brillouin imaging is its general inability to deal with noise.

\begin{figure}[H]
	\centering
	\includegraphics[width=\textwidth]{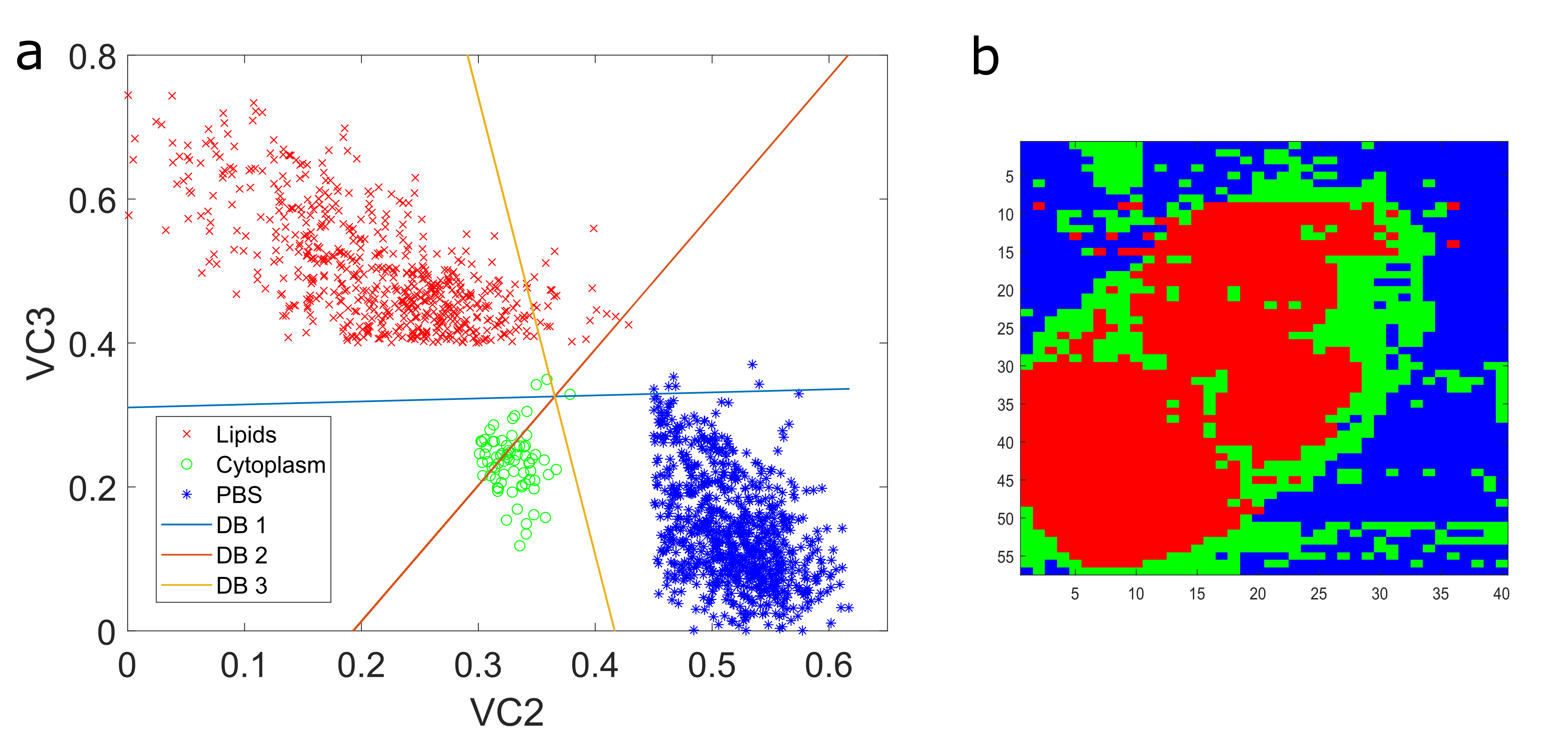}
	\caption{\textbf{Supervised classification of an unseen biological dataset. a} Scatter plot representation of the training dataset, using the scores of the second and third vertex components of the data. Three decision boundaries were identified using LDA, DB1 for lipids and cytoplasm, DB2 for lipids and PBS and DB3 for cytoplasm and PBS. \textbf{b} RGB image showing the result of performing spectral classification on the data obtained from the cell in Figure \ref{fig:cell2}a, the same colour scheme applies.}
	\label{fig:cell3}
\end{figure}

\begin{figure}[h!]
	\centering
	\includegraphics[width=\textwidth]{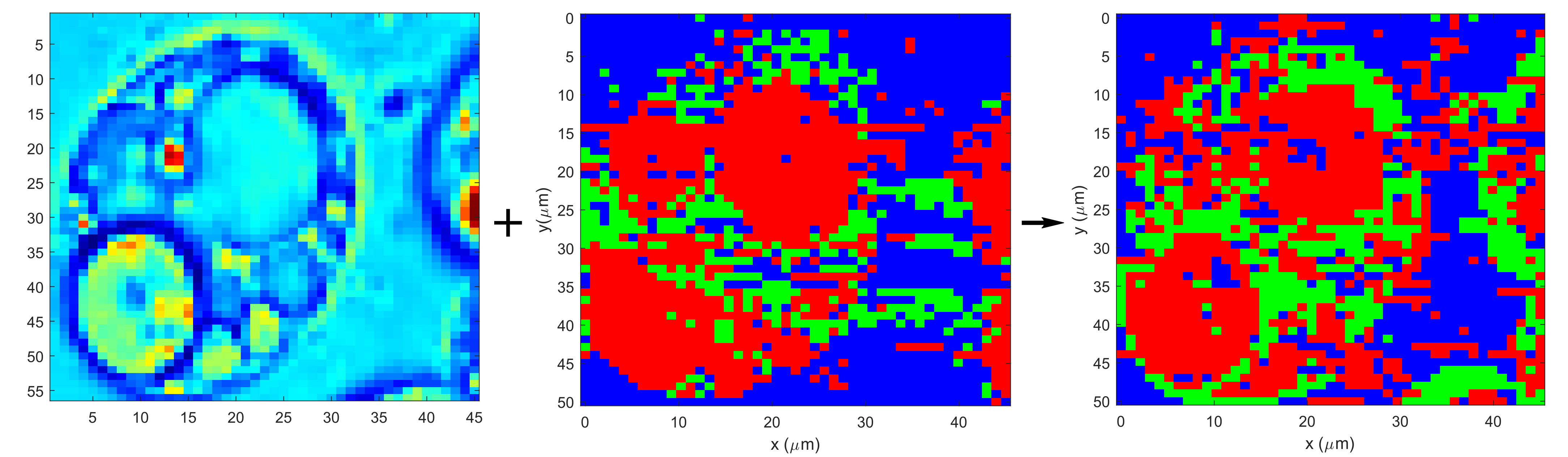}
	\caption{\textbf{Illustration of the process and results of training HCA with intensity information.} From left to right, the pixelated intensity image obtained from modifying image 4a, RGB image of the cell obtained by normal HCA and RGB image of the same cell obtained by trained HCA.}
	\label{fig:cell4}
\end{figure}

To compare with a supervised approach, %The images in Figures \textcolor{red}{1} and \textcolor{red}{2} were designated as the training and test datasets respectively.%
a LDA workflow was applied to first train a predictive model using the data obtained from another adipocyte cell as training data. The performance of this model was then tested on the same cell presented in Figure \ref{fig:cell2}, which is a set of independent, test data from the model's perspective. Independent labelling of the training dataset was achieved by taking advantage of the dimensionality reduction enabled by VCA, which was used to identify the distinct spectral components present. By studying the component scores plot, maps of the three materials present (i.e. lipids, cytoplasm and PBS) were then created by thresholding the second and third vertex components (VCs), which qualitatively were observed to be related to PBS and lipids respectively (see Suppplementary materials). The labelled scores were then used to train a LDA classifier (Figure \ref{fig:cell3}a) to divide the data into three classes, which was then applied to the second and third VCs of the test dataset, obtained by projecting the data onto the same vertex components that had been previously identified from the training dataset. The results of the supervised classification are presented in Figure \ref{fig:cell3}b, where a RGB image similar to those obtained from unsupervised clustering was also generated to visualise the biological features. It can be seen that while the supervised approach was generally better at identifying the cellular features (i.e. lipids and cytoplasm), it also suffers from background variability in the data, in this case caused by stitching. Quantitative evaluation of the performance, such as calculation of sensitivity and specificity were not attempted due to the unclear definition of the ground truth. For future implementations, sufficient pre-processing of the data is advisable in order to fully realise the power of supervised learning that was qualitatively demonstrated here, which underpins the use of BI as a diagnostic tool.
\noindent To address the issue of noise and background variability, \emph{a priori} information from the intensity BF image was incorporated to train the HCA process using our proposed semi-supervised method, with a weighting factor of $w=0.2$, the resulting images are shown in Figure \ref{fig:cell4}.
Even with $w=0.2$, it can be seen that clusters have benefited from cleaner segmentation, thus decreased the occurrence of fuzzy pixels. To verify this quantitatively, average spectral parameters for the three regions were first extracted by selecting relevant ROIs in the image and finding their mean values through simple line fitting, which are then treated as the ground truth. It was found that the average spectra of the relevant clusters produced Brillouin shift values that were $1-2\%$ closer to these ground truth values \cite{wu2020}, which can be as much as $100$ MHz. This does not suggest improvement with confidence, however, as the ground truth ROIs are chosen somewhat subjectively due to the lack of any objective measure in this case. Any improvement should be quantified in terms of classification accuracy in the future, ideally with a labelled training set. At the same time, any further increase in the weighting will cause the image to be increasingly dominated by the intensity contrast, albeit becoming less noisy. If the \emph{a priori} information matches well with the actual hyperspectral data, this might not be an issue as more clusters simply have to be averaged for one region of contrast. While an intensity-based model may not be the ideal choice for training in all cases, it is possible to switch to a more Brillouin-relevant model, for example a biphasic \cite{BRI_kathy2018} or poroelastic model \cite{Margueritat2019}.  

\section{Conclusion}
In this work, we have presented multivariate algorithms, both supervised and unsupervised for the enhanced analysis of hyperspectral data acquired by Brillouin imaging. Specifically, we have demonstrated in Brillouin data of a phantom and cells, how different methods can be applied independently and in conjunction to perform spectral unmixing, classification and segmentation. From our results, we have shown that the unmixing of spectral data produces images that are high in contrast owing to deciphering of sub-pixel information contained in a spatially impure voxel and the general noise resilience of algorithms. When applying traditional line fitting on the same data, the resultant images were either missing key features due to the averaging of spectral information or produced erroneous results due to under-fitting. Apart from producing abundance images that bear striking resemblance to the reference microscopy images, the algorithms utilised in this work are also highly advantageous in terms of time efficiency, having reconstructed images of biological samples in a time approximately $400$ times faster than line fitting. The corresponding spectral analysis is also more efficient as only the endmember spectra are studied, thus decreasing the computational intensity whilst retaining peak shift values that are comparable to the values obtained from fitting the entire dataset. The spectral shape and hence linewidth information from the endmember spectra, however, require closer attention during unmixing. The distortion in the spectral profile is explainable by the underlying assumption of the algorithm, that the numerical implementation simply has to predict the most likely spectra that explains a linear mixing model, which may not exist in the actual data. Even though the robustness enabled by this method is desirable in complex datasets, in order to retain the information in peak widths, future implementations would benefit from choosing a larger window in the spectral domain for prediction, or alternatively by increasing the sampling rate of peaks. The response function of the instrument should also be included in the algorithm if the true values are needed. Lastly, while a linear mixing model has produced sufficient performance thus far, some biological experiments are likely to yield data that are based on an intimate mixture. For example, if a single voxel contains spectral information from a stratified structure, the abundance of the hidden endmembers would then depend on that of endmembers on top, therefore requires non-linear unmixing, which is far less developed and remains a key area of research for future work \cite{Dobigeon2014}. 

To the best of our knowledge, we have also demonstrated for the first time feature identification in Brillouin images. Namely, the pixels can be classified by a pre-trained model in a supervised fashion or segmented by their intrinsic inter-pixel similarities, defined by a unsupervised measure. While unsupervised techniques, represented by cluster analysis, provide an avenue towards online, real-time data analysis in the future, the limitation of noise is detrimental and will have to be overcome for clinical applications. In this vein, we proposed optimisation by training the algorithm with \emph{a priori} intensity information, which was able to form images with better quality to a certain extent. Excessive training, however, would also overwrite the mechanical information in the Brillouin images with intensity information. Future improvement would require far more training data than what is currently available, as well as better, deeper understanding of the Brillouin interaction. Finally, it should be noted that the scope of supervised or unsupervised spectral labelling extends beyond a single image, as larger datasets consisting of multiple images, obtained from different samples can also be analysed, with the aid of other machine learning tools that were not explicitly covered in this work, such as artificial neural networks (ANN) \cite{Lundervold2019}. In conclusion, further development of these algorithms will be key in the construction of a standardised workflow, from pre-processing to post-processing, for the future of Brillouin hyperspectral imaging.

\section*{Funding}
UK Engineering and Physical Sciences Research Council.

\section*{Disclosures}
\noindent The authors declare that there are no conflicts of interest.	

\bibliographystyle{ieeetr}
\bibliography{library}

\pagebreak
\appendix
%\counterwithin{figure}{section}
%
\setcounter{figure}{0}
\renewcommand{\thefigure}{A\arabic{figure}}

\end{document}